# High Performance I/O For Large Scale Deep Learning

Alex Aizman
*NVIDIA*
Santa Clara, CA, USA
aaizman@nvidia.com

Gavin Maltby
*NVIDIA*
Santa Clara, CA, USA
gmaltby@nvidia.com

Thomas Breuel
*NVIDIA*
Santa Clara, CA, USA
tbreuel@nvidia.com

*Abstract*—Training deep learning (DL) models on petascale datasets is essential for achieving competitive and state-of-the-art performance in applications such as speech, video analytics, and object recognition. However, existing distributed filesystems were not developed for the access patterns and usability requirements of DL jobs. In this paper, we describe AIStore, a highly scalable, easy-to-deploy storage system, and WebDataset, a standards-based storage format and library that permits efficient access to very large datasets. We compare system performance experimentally using image classification workloads and storing training data on a variety of backends, including local SSDs, single-node NFS, and two identical bare-metal clusters: HDFS and AIStore.

## I. INTRODUCTION

Training deep learning models on petascale datasets is essential for achieving competitive and state-of-the-art performance in applications such as speech, video analytics, and object recognition [1]. However, existing distributed filesystems were not developed for the access patterns and requirements of deep learning jobs. As datasets grow to scales that strain the capabilities of the conventional storage solutions, deep learning workloads require a new generation of distributed storage systems.

Deep learning methods usually iterate repeatedly through random permutations of a training dataset. Besides, they also often require some kind of map-reduce style preprocessing, whereby shuffling and parallel processing requires random access. On the other hand, high data rates are typically associated with a sequential large-block I/O. To satisfy these competing demands, researchers often transform deep learning datasets to contain collections of shards that, in turn, combine or comprise original file data. Such sharded dataset storage requires certain conventions and software components: a storage format, a client library for reading it, and a server for serving the shards.

Google uses TFRecord/tf.Example as its storage format, the TensorFlow library for reading it, and GFS as the file server component. Hadoop uses Apache Parquet and HDFS, together with Java-based client and server libraries. Google's solution is partially proprietary (GFS is not available, and there are not many tools for processing TFRecord/tf.Example files), and the Hadoop toolchain is not well suited to deep learning applications, in part because of its reliance on the Java runtime and its focus on processing data structures rather than files.

## II. BACKGROUND

At a high level, training a DL model on a given dataset consists of the following 3 (three) fundamental steps:

(1) **randomly shuffle** the dataset;
(2) **sequentially iterate** through the shuffled dataset:
   a. read a **batch** of samples;
   b. deliver decoded (augmented) batch to a given DL job that runs on a given GPU core;
(3) once the entire dataset is traversed:
   a. check training objective (accuracy); update gradients;
   b. goto (1) if needed; otherwise, stop.

Further, and as a matter of terminology, a single DL traversal through the entire dataset is often referred to as one *epoch*; each epoch is preceded by an implicit or explicit reshuffling of the training dataset.

Note that, even though simplified and high-level, the sequence above **emphasizes** important aspects shared by all deep learning workloads: random shuffle followed by sequential traversal of the (shuffled) dataset, operation on a batch-of-samples basis, whereby a given computing GPU or CPU core computes over one batch at a time.

From the software engineering perspective, this commonality immediately translates as (the need for) framework-agnostic solution to optimize DL workloads. In this study, we present our solution that we also benchmark using PyTorch [2] (with future plans to benchmark other frameworks as well, TensorFlow in the first place).

## III. DISTRIBUTED STORAGE: STATE OF THE ART

Generally, the storage technology that is currently utilized by (and for) Big Data machine learning falls into the following 4 broad categories:

- Parallel POSIX filesystems
- Object storages
- Google GFS and Amazon S3
- HDFS

TensorFlow [3] explicitly supports all 4 storages via its file system interface abstraction:

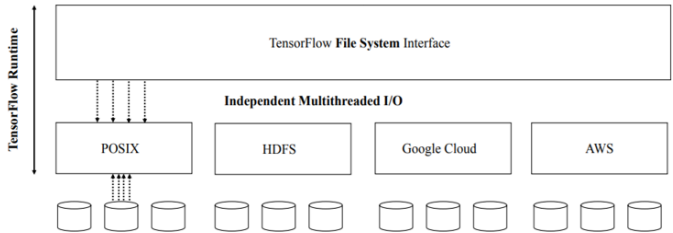

Fig. 1. TensorFlow File System Interface, courtesy of [4].

The same is largely true for all popular DL frameworks [5, 6, 7], although there are, inevitably, nuances. PyTorch, for instance, does not have a dedicated HDFS DataLoader, but that simply means that HDFS can be accessed via custom DataLoader that would utilize existing Python ⇔ HDFS bindings.

In conclusion, several distributed storage solutions for Big Data processing have been developed – the list includes Google's GFS and Hadoop's HDFS. Such systems generally provide high-throughput I/O at the tradeoff of higher latencies. They are, therefore, usually deployed with record-sequential (aka, sharded key-value) storage formats such as TFRecord (TensorFlow) and Apache Parquet (Hadoop). While such systems can theoretically meet the I/O requirements of large-scale compute jobs, there are practical problems and limitations when it comes to their adoption as storage solutions for distributed deep learning. Among those are lack of tools, custom protocols, difficulties in deployment, limitations in using existing serialization formats, and, perhaps most importantly, substantial storage-specific changes to data processing pipelines. Thus, the requirements for a large-scale deep-learning storage solution include use of standard, widely supported protocols and formats, easy migration of existing DL models and datasets, scalability that allows storage to be accessed at speeds close to hardware capabilities, easy setup and configuration, predictable performance, compatibility with tools for Python and command-line ETL, and finally, easy integration with Kubernetes.

Our solution that meets these requirements utilizes AIStore [8] and open-source Python libraries.

IV. AISTORE

By design, AIStore (or AIS) provides an infinitely scalable namespace over arbitrary numbers of disks (SSDs and/or HDDs) while the conventional metadata-server related bottlenecks are eliminated by having data flowing directly between compute clients and clustered storage targets.

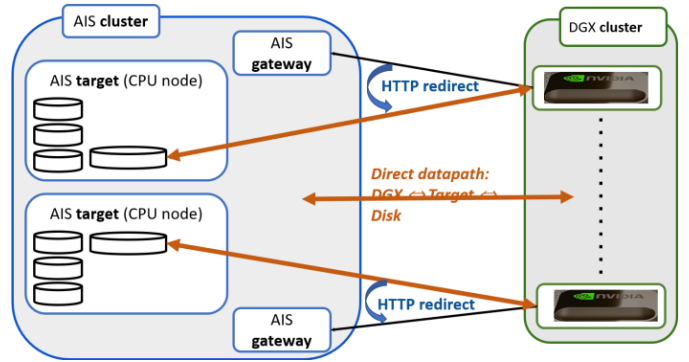

Fig. 2. AIStore: direct datapath, linear scale-out.

At its core, AIS is a lightweight, minimalistic (scale-out) object store providing an S3-like RESTful interface: read/write access to user data is achieved through simple HTTP GET and PUT operations, respectively. High I/O performance is achieved, in part, by relying on HTTP redirects: a dataset may consist of billions of objects, but when a client requests an object, the client retrieves that object via a direct connection to the storage server holding that object.

On the backend, AIS natively integrates with Amazon S3, Google Cloud Storage (GCS), and other S3 or GCS-compliant object storages. User can explicitly prefetch or download all, or selected, objects from S3 and/or GCS buckets, or load requested objects on-demand, thus effectively utilizing AIStore as a high-speed caching tier capable to scale-out with no limitations.

AIS is written from scratch in Go, fully open-sourced, and runs on any commodity hardware. Deployment options are practically unlimited and include a spectrum with bare-metal Kubernetes, on the one hand, and a single Linux host, on the other ([9]). In fact, Kubernetes proved to be quite instrumental when there's an immediate need to deploy AIStore clusters ad hoc as a large-scale high-performance tier between traditional storage systems holding cold data and deep learning jobs. In addition to end-to-end data protection, n-way mirroring, and m/k erasure coding on a per bucket (per dataset) basis, AIStore integrates MapReduce extension to reshard distributed datasets given user-defined sorting order and shard size – the two parameters that are crucially important for the subsequent training.

V. WEBDATASET

To speed the adoption of sharded sequential storage for deep learning, we have defined a simple storage convention for existing file-based datasets based on existing formats. In particular, WebDataset [10] datasets are represented as standard POSIX tar files in which all the files that comprise a training sample are stored adjacent to each other in the tar archive. Such tar archives can be created with simple UNIX commands. Not only can they be used for training, but they are also a fully archival representation of the original file data.

To support WebDataset, we have created a companion Python library that provides a drop-in replacement for the built-in PyTorch Dataset class and allows existing filesystem-based code to be converted to access record-sequential (aka sharded key-value) storage with minimal changes. In particular, decompression, augmentation, and parallelization continue to function exactly as in the file-based pipelines. The WebDataset library can read from any input stream, including local files, web servers, and cloud storage servers.

## VI. SMALL-FILE PROBLEM

The proverbial "small-file problem" affects all storage solutions and is well-known and well-documented – see, for instance, [11, 12, 13]. On the other hand, large scale DL datasets often comprise hundreds of millions, or even billions, of samples totaling massive volumes of, effectively, small files.

There are numerous ways, supported by many familiar archival tools, to transform a small-file dataset into a dataset of *larger shards*. In this study, we used AIStore-integrated dSort to reshard inflated ImageNet [14] to shards between 128MB and 1GB sizes. AIStore dSort creates shards in parallel by all storage nodes, thus optimizing dataset transformation time. But the bigger point is that in the end, even though (re)sharding operation is optional, it is often more optimal to spend some extra time on it – the point that becomes progressively more pronounced when datasets grow in size to contain millions of small files.

The second point – scalable storage access – is also crucial. By design, AIStore supports direct client ⇨ storage target data flow, whereby a READ or WRITE request gets quickly HTTP-redirected to the designated storage server (aka AIS target). The redirection is accomplished by any number of deployed AIStore gateways (aka AIS proxies) that are stateless, extremely lightweight, and can run anywhere. In particular, the (supported) option to run AIStore gateway on each storage-accessing client allows optimizing redirection time down to microseconds.

Once redirected, the data flows directly between a client and a clustered storage node. By design, AIStore gateways never "see" a byte of user data while at the same time providing a unified global namespace for the combined terabytes and petabytes of user content.

## VII. SHARDS AND RECORDS

Large scale DL datasets may comprise hundreds of millions, or billions, of samples totaling massive volumes of data. Samples are often small in size. Even if the original raw data files are large (e.g., video), they are often split out and prepared as much smaller samples for training data. Hence, the problem of small random reads.

At the time of this writing, 4KB random read throughput for SSDs ranges anywhere between 200MB/s and 900MB/s [15], thus at the low end matching sequential read performance of a stock enterprise HDD. Given the common characteristics of deep learning workloads, one can only draw the conclusion that performance and scale (and, in particular, performance at scale) requires optimized random reading at sizes orders of magnitude greater than 4KB. Notwithstanding, DL datasets that consist of unaggregated individual samples (aka small files) are quite common today - the observation that leads to another (quite common) course of action where a serious time and effort is spent on optimizing deep learning pipeline for those small-file datasets (see, e.g., [4]).

In this paper, we strongly advocate the preprocessing step of aggregating *smaller* samples into *bigger* shards. This can be done using, for instance, any of the available and familiar archival tools, e.g., the ubiquitous GNU tar. With AIS, user-defined pre-sharding is also automated and fully offloaded to a storage cluster, with an immediate benefit of large-size reads, that further translate into "extracting" vendor-documented throughputs from clustered drives: HDDs and SSDs.

What could then be the best, toolchain and IO friendly, sharding format? The one well-known and TensorFlow-native is, certainly, TFRecord/tf.Example. There are tradeoffs, though, associated with using it, especially outside TensorFlow, and in particular for deep learning (with its media-rich input formats); however, full discussions of those tradeoffs would take us out of the scope of this paper. Suffice it to say, we strongly advocate an open-source, open-format, ubiquitous, and widely-supported: GNU tar.

### A. Definitions

**Shard:**
- protected unit of storage
- unit of computation
- contains files or *Records*
- *may be* a GNU .tar/.tar.gz

**Record:**
- abstracts objects files with the same basename wo/ ext.; is indivisible
- e.g., records A and B:
  - [A.png, A.cls, A.json]
  - [B.png, B.cls, B.json]

**Sharded Dataset:**
a unit of applying storage policies (e.g., num copies)
transformed DS contains sharded *records* from the source

Fig. 3. A *shard*, a *record*, and a *sharded dataset*.

In summary, Big Data datasets – in particular, datasets used for DL training – can be stored in shards formatted as tar files that contain *records* that, in turn, comprise original files along with their *associated* metadata. For deep learning, the most natural and effective way to express the *record association* appears to be filename without extension (Fig. 3).

### B. GNU tar

Humble tarfile, as we have rediscovered, appears to be a simple-albeit-straightforward and, ultimately, very powerful tool to convert any given small-file dataset into a sharded (Fig. 3) one.

The decision to use tar immediately benefits from universal support by virtually all languages and toolchains. It is extremely easy to supplement existing libraries with a set of DL-specific tools that use tar - see, for instance, our open-source *tarproc* utilities for Python [16].

The fact that tar (and tar.gz) simultaneously works as a data archive providing additional data protection, and an optimized data source – cannot be underestimated.

## VIII. THE SOLUTION

In summary, the core ideas that we propose and implement to support the new DL pipeline:

1) large reads;
2) highly scalable storage access protocol;
3) independently scalable pipeline stages: I/O, decoding, augmentation, deep learning;
4) easy ability to assign any component to any k8s (compute or storage) node.

The system combines agile easy-to-deploy scalable open storage that runs on any commodity hardware, and an integrated data processing pipeline that can flexibly execute on storage and/or compute nodes while posting ready-to-compute tensors via RDMA directly into GPU memory (Fig. 4):

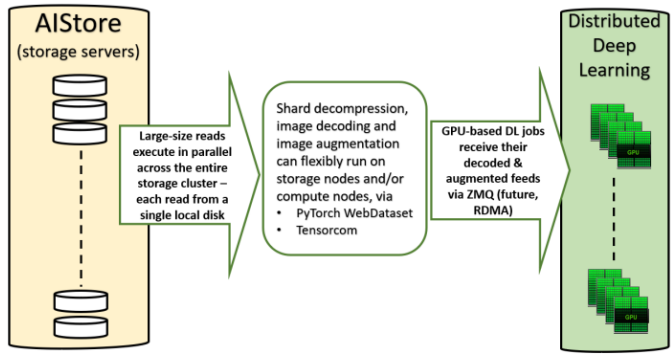

Fig. 4. Integrated data processing pipeline.

Overall, the system that we propose combines agile, easy-to-deploy-anywhere scalable open storage and an integrated data processing pipeline that can execute on storage and/or compute nodes while posting ready-to-compute tensors into GPU memory (Fig. 1).

## IX. BENCHMARKS

### A. Methodology

There is currently a lack of established/standardized deep learning benchmarks (see [17]) and particularly, in our case, benchmarks that run distributed deep learning (DDL) workloads while isolating the contribution of storage I/O to DDL performance and the respective bottlenecks. It was tempting to fall back to artificial benchmarks of synthetic "DDL like" loads that try to demonstrate throughput and latency. But given the complexity of the DDL data pipeline, it would be difficult to establish (with a suitable degree of confidence) that the performance so demonstrated really has any relevance to a real DDL application.

Rather than investing time in attempting a synthetic benchmark that would likely still leave open questions of validity and relevance (never mind the lack of comparison with existing storage solutions), we decided to benchmark the performance of end-to-end training and inference in a particular DL framework with a fixed DL model. In the end, we selected PyTorch and ResNet-50, respectively, for their familiarity, and for the ready availability of many reported results ([17]).

It is important to recognize that the aim/expectation is not to equal or better the DDL performance achieved when working with (usually small) datasets served from fast local SSD storage or from DRAM (cached filesystem, or more specialized such as LMDB). Instead, our objective is to demonstrate that those established performance levels can be reasonably matched and sustained as dataset size grows well beyond the practical/economic capacity of a single node.

In deciding to benchmark end-to-end DL for various storage backends, the metric of interest is – how quickly the training/evaluation loop iterates and consumes data from the DataLoader pipeline. A fixed batch size of 256 was used, and 200 iterations of the loop timed (51200 images per GPU) with the aggregated distributed result expressed in MB/s based on the average image file size for the dataset. All testing used FP16 models, the NCCL [18] distributed backend, and 5 DataLoader workers per GPU (separate testing confirmed these defaults were optimal for all DataLoader backends).

### B. Setup

The hardware that we had at our disposal included 8 NVIDIA® DGX-1™ stations, 12 storage servers, and a 100GbE Arista switch:

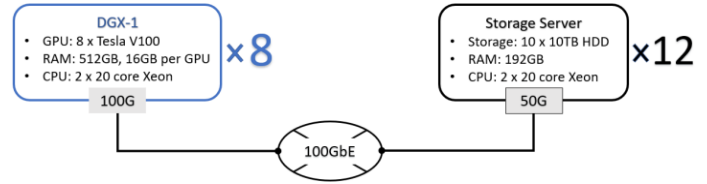

Fig. 5. Hardware setup.

Based on this, we built and then benchmarked several single-node and clustered configurations described below in greater detail. The software that we have used included:

- Ubuntu 18.04 LTS
- AIStore v2.5
- Kubernetes v1.14.1
- HDFS v3.2.0

and a recently-released PyTorch 1.2 built from source with PyTorch NGC container version 19.07-py [19].

## X. PERFORMANCE: DISTRIBUTED DEEP LEARNING

We compared DL performance using PyTorch-based ResNet-50 models and different storage backends used to store datasets of different sizes. All datasets were derived from ImageNet [14] by uniformly duplicating existing samples under randomly generated file names. On the client side, we used WebDataset [10] that provides a drop-in replacement for the namesake PyTorch Dataset class and allows existing filesystem-based code to seamlessly work with tar shards. Overall, our objective was to measure the *end-to-end* performance of the whole (compute + storage) system and to avoid common DL-benchmarking pitfalls [18].

The table below summarizes end-to-end deep learning benchmarks in terms of their respective storage backends:

TABLE I.
END-TO-END RESNET50 TRAINING: DATASETS AND STORAGE BACKENDS

| Benchmark "label" used in the plots below | Dataset | Objectives |
|---|---|---|
| SSD; "ssd" | 7TB-file | Establish perf. baseline at single compute-node capacity limit |
| Standalone Ubuntu 18.04 kernel NFS with 10 HDDs in a single LVM 256KB stripe; "nfs" | 7TB-file | How does NFS cope with an uncacheable dataset? |
| HDFS 12 node, 120 HDD; "hdfs12" | 7TB-file 7TB shards 65TB shards | State of the art distributed filesystem (HDFS); non-sharded and sharded datasets |
| AIS 12 node, 120 HDD; "ais12" | 7TB-shards 85TB-shards | AIStore, sharded input, datasets at single node limit and well beyond single node capacity |
| AIS 12 node, 120HDD; "ais12" | 238TB-shards | Demonstrate consistent at-scale performance |

Results in Fig. 6 show that our solution, when deployed on commodity hardware and rotational drives, delivers local SSD-like performance. The legend is documented in the summary table above. For instance, "hdfs12-7TB-shards-webdataset" indicates ImageNet *inflated* to 7TB of total size, sharded into 1GB size shards, and serviced by a 12-node HDFS cluster via WebDataset.

The "ssd-7TB-file-pytorch" and the "nfs-7TB-file-pytorch" plots correspond to local DGX SSDs and a single-server NFS, respectively, and are included as a *non-distributed baseline* for further comparison of per-GPU and per-disk performance numbers.

Notice that AIStore and HDFS show a similar linear scale – the fact that is likely attributable to the (compute) client side being the bottleneck (Fig. 6). This observation was confirmed in all our DDL training and inference benchmarks (omitted here due to space constraints).

Consider also Fig. 7 results for sharded input data served by two Kubernetes clusters (utilizing the same exact hardware – see Fig. 5):

- AIStore – **238TB** *inflated* ImageNet dataset

vs

- HDFS – 7TB *inflated* ImageNet dataset

.

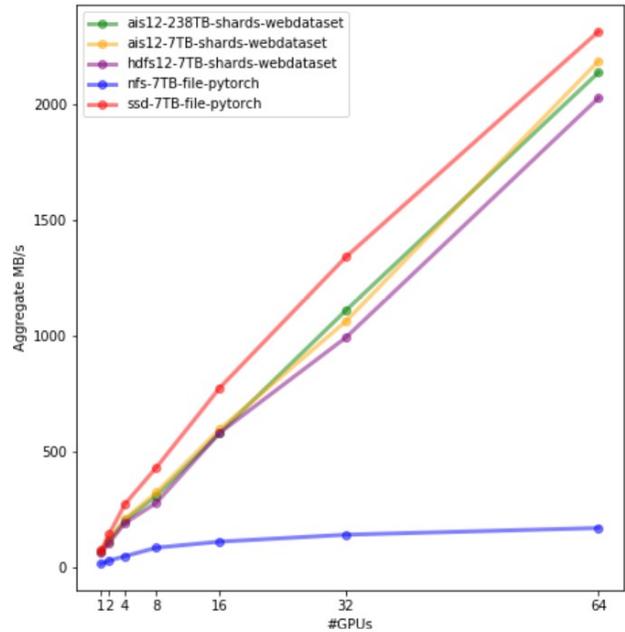

Fig. 6. ResNet-50 training on a variety of *inflated* ImageNet datasets and storage backends (for labels see Table I).

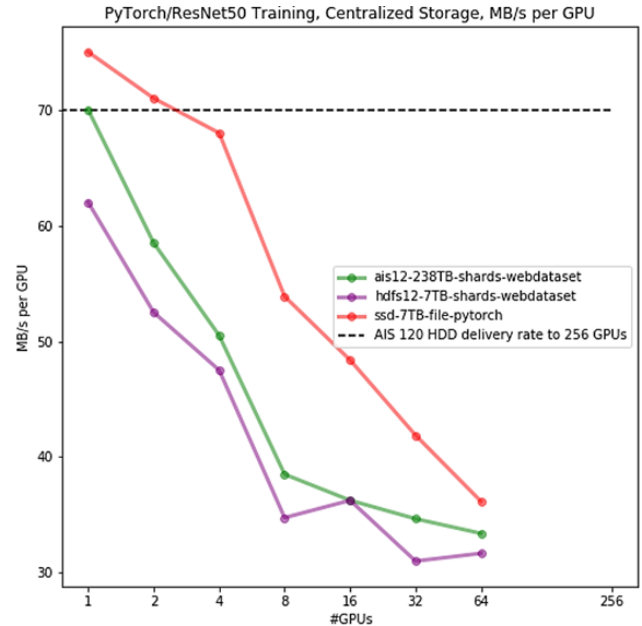

Fig. 7. ResNet-50 training: throughput per-GPU.

This result again demonstrates the benefits of sharding, on the one hand, and the limits of GPU ResNet-50/PyTorch training, on the other. It becomes clear that, given DGX-1 on-SSD training rate 54MB/s, one HDD can "feed" approximately 2.5 GPUs out of 8 (eight) GPUs in the DGX. Normalizing the aggregate data rate across all GPUs to a rate per GPU demonstrates that, while PyTorch/ResNet-50 initially demands ~60-70MB/s per GPU, the demand drops to more like 40-50MB/s per GPU beyond a total of 8 GPUs. Provided the storage backend scales appropriately, that demand level suggests each HDD can feed approximately 3 GPUs in a distributed deep learning.

## XI. PERFORMANCE: MAXIMUM DATA DELIVERY RATE

WebDataset and, of course, AIS are not tied to PyTorch itself, and so the next natural question is how the system will perform with other, potentially more optimized, deep learning frameworks? Stated differently, how many deep learning clients can the given storage cluster configuration support at a reasonable data rate?

To this end, the following benchmark (Fig. 8) runs on an 85TB dataset comprising 68,000 x 1.25GB input shards, each shard containing, on average, 8,600 input images for a total of 588 million images. HDFS utilizes its default 128MB block size while AIS is configured to store entire shards on the same 120 HDDs (of 12 clustered nodes). Data redundancy-wise, both HDFS and AIS maintain 3 replicas of each of the 68K shards.

Note that this setup *favors* HDFS (vs. AIS) to a degree: a given 1.25GB shard is striped across up to 10 HDFS drives and loaded in parallel. The test load simulates DataLoader worker processes on compute (aka GPU) nodes: 5 workers per GPU. A 40-worker load is run on a single GPU node, then 80 workers on 2 nodes, and so on up 280 workers over 7 compute nodes and a final additional result of 8 nodes with 120 and with 360 workers each. The benchmark selects input shards at random, reads an entire shard, and discards the read data (Fig. 8).

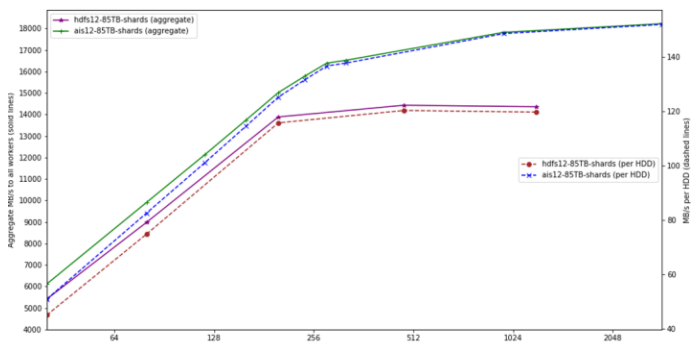

Fig. 8. Maximum data delivery rate: aggregate throughput for all workers (**left axis**) and the throughput per HDD (**right axis**).

## XII. DISCUSSION

Overall, in end-to-end DL testing with sharded input (as per Table I), the performance of the 12-node (120 HDD) AIS and HDFS clusters is very similar. This is a good result for AIS as HDFS is a mature distributed filesystem. The sharded nature of the input data helps avoid HDFS NameNode scalability issues [4]; HDFS, however, did not perform well in our benchmarks with the original (non-sharded and non-inflated) ImageNet [14] that contains, on average, 140KB size images, and, therefore, can be considered a *many-small-files* use case. With no centralized NameNode equivalent, AIStore scales better than HDFS as the number of clients increases – our future work will further assess and test this observation.

In tests of maximum data delivery rate (Fig. 7), AIStore delivers 18GB/s aggregated throughput, or 150MB/s per each of the 120 hard drives – effectively, a hardware-imposed limit. HDFS, on the other hand, falls below AIS, with the gap widening as the number of DataLoader workers grows. The biggest single factor in that delta is most likely client overhead: each HDFS client requires a JVM instance, as in:

Python ⇨ DataLoader ⇨ PyArrow [20] ⇨ libhdfs [21] ⇨ JVM

Starting up a DL load with HDFS proved to be somewhat difficult as it entails concurrent listings of HDFS directory by all running DataLoaders, which in turn requires tuning of the respective JVM heap sizes. The problem is illustrated in Fig. 2, where we encountered a limit of no more than 40 workers per GPU node. There must be, however, alternative ways to run HDFS in DL environments; it is also possible that we have overlooked HDFS-specific performance tuning. For any/all of these reasons, our future work will certainly include more (and larger-scale) AIStore ⇔ HDFS comparisons.